\documentclass[10pt]{article}
\usepackage{makeidx,amsmath,amsfonts,latexsym,subfigure,color,graphicx}
\usepackage{natbib}
\usepackage{setspace}
\usepackage[utf8]{inputenc}
\usepackage[affil-it]{authblk}

\newcommand{\E}{{\mathbb E}}

\newcommand{\Pb}{{\mathbb P}}

\begin{document}

\title{Fermi's paradox, 
extraterrestrial life and the future of humanity: a Bayesian analysis}

\author{Vilhelm Verendel
  \thanks{Corresponding author. Electronic address: \texttt{vive@chalmers.se}; Adress: Physical Resource Theory, Department of Energy and
Environment, Chalmers University of Technology, 412 96, Gothenburg,
Sweden. Telephone: +46 31 772 3107. Fax: N/A. }}
\affil{Department of Energy and Environment, Chalmers University of Technology, 412 96, Gothenburg, Sweden}

\author{Olle Häggström
  \thanks{Electronic address: \texttt{olleh@chalmers.se}}}
\affil{Department of Mathematics, Chalmers University of Technology, 412 96, Gothenburg, Sweden}

\maketitle

\newpage
\doublespacing

\begin{center}Abstract\end{center}
\bigskip\noindent
{\bf
The Great Filter interpretation of Fermi's great silence
asserts that $Npq$ is not a very large number, where $N$ is the number
of potentially life-supporting planets in the observable universe, $p$
is the probability that a randomly chosen such planet develops
intelligent life to the level of present-day human civilization, and
$q$ is the conditional probability that it then goes on to develop a
technological supercivilization visible all over the observable
universe. Evidence suggests that $N$ is huge, which implies that $pq$
is very  small. Hanson (1998) and Bostrom (2008) have argued that the  
discovery of extraterrestrial life would point towards $p$ not being
small and therefore a very small $q$, which can be seen as bad news for
humanity's prospects of colonizing the universe. Here we investigate
whether a Bayesian analysis supports their argument, and the answer
turns out to depend critically on the choice of prior distribution.
}

\newpage
\section{Introduction}
\noindent 
Recent discoveries such as Earth-like planets in the habitable zone
around another star \citep{Quintana14,cassan2012one} could plausibly
increase expectations to find life elsewhere and possibly under
similar conditions as on planet Earth. However, in contrast to popular
opinion, Hanson (1998) and Bostrom (2008)  have suggested that a
future discovery of extraterrestrial life would be bad news for
humanity.  The reasoning behind this view rests on Hanson's
so-called Great Filter formalism \citep{Aldous,Hanson}. The number $N$
of potentially life-supporting planets is huge (perhaps around
$10^{22}$) \citep{cassan2012one}. Take $p$ to be the probability that a randomly chosen such
planet develops life and a technological civilization on the level of
humanity today,  and $q$ to be the probability that a randomly chosen
planet  having reached that level goes on to form a
super-technological civilization, defined as one that is noticeable to
astronomers everywhere in the observable universe.  Then the expected
number of such super-technological civilizations is $Npq$. The
great silence of the Fermi Paradox \citep{Webb,C09} 
indicates that the actual number of such super-technological
civilizations is $0$, suggesting that the order of magnitude of $Npq$
is at most $1$.  The fact that there is an astronomical number of
exoplanets would imply that $pq$ is very small, but discovery of
extraterrestrial life would be evidence that $p$ is not very small, in
which case $q$ must be very small, which in turn suggests that
humanity's prospects of eventually reaching the maturity of a
super-technological civilizations are bleak. 
In the present paper we investigate whether or not a rigorous
statistical analysis supports this.

The issue of what the discovery of extraterrestrial life would mean to
humanity has a certain timeless flavor to it. Yet, it is perhaps more
pressing now than ever, for two reasons. First, it has become increasingly
clear that humanity in the coming century or so faces a number of
non-negligible risks threatening our very existence \citep{BC,PA,B14,Martin06,Rees03}; a
very small value of $q$ might be an indication that the situation requires
even more care than we have hitherto realized. Second, the discovery of
exoplanets, many of which seem to be potentially life-supporting,
is proceeding at a rapid pace \citep{PHM}, and perhaps 
we will soon attain the ability to detect definite
signs of life out there \citep{BS}. 

We write
\begin{equation} \label{eq:GreatFilter}
Npq \not\gg 1 \, ,
\end{equation}
for the claim that $Npq$ has order of magnitude at most $1$.
This relation is called the Great Filter, as it says that all or almost
all planets are filtered out somewhere on the path from its genesis to
the emergence of a super-technological civilization. The Great Filter
has a predecessor in the Drake equation for the number of
civilizations active in radio astronomy in the Milky Way, which
was formulated by Frank Drake in 1961, and which
since then has enjoyed an iconic status
in the search for extraterrestrial intelligence \citep{Webb}.
The Great Filter differs from the Drake equation, not only in a
choice of factorization
that zooms in on the future prospects for present-day humanity, but also
in a number of other aspects, including lifting the focus from the 
Milky Way to the entire visible universe, a move that is motivated by
the apparent in-principle feasibility of both intergalactic 
colonization (over long time scales) and the construction of engineering
structures visible over intergalactic distances
\citep{AS13,GWMPSM,H16}. While it is conceivable that there could be
very advanced civilizations that are for some reason invisible
to us, basic evolutionary arguments and the astronomical time scale of
the universe suggest that if advanced technological civilizations
were common, some of them would have been observable here long ago
\citep{Hart75,Hanson,H16}. Again, this suggest that $pq$ has to be really
small.

It seems to us worthwile to investigate whether the arguments
of Hanson (1998) and Bostrom (2008) 
on the discovery of extraterrestrial life as bad news to humanity
survive a more rigorous statistical analysis. In the choice between a
frequentist statistical formalism and a Bayesian one, we opt for the
latter. A general reason in favor of a Bayesian framwork is that it
is the only one that can produce probabilistic statements about the
whereabouts of the parameter values of interest. A more specific
reason is that the philosophy behind frequentist statistics, with its
emphasis on independent repetitions of experiments
\citep{CH,Salsburg}, seems ill-suited to the present context, as we
observe Fermi's great silence once and for all, so that the idea of
independently repeating the experiment borders on the nonsensical. A
downside of the Bayesian approach is that (as we shall see) the
results may depend on the choice of prior distribution, especially so
when data are sparse. 

\section{Analysis}
To accomodate the possibility of observing {\em primitive} life on another
planet, we choose to factorize $p$ in the Great Filter equation 
(\ref{eq:GreatFilter}) one
step further as $p=rs$, where $r$ is the probability that
a randomly chosen planet with potential for supporting life 
does develop life to the level of (say) amoebas, and $s$ is the probability
that such a planet goes on to the level of persent-day human 
technological civilization, conditional on having reached the amoeba level.
Thus, (\ref{eq:GreatFilter}) turns into
\begin{equation} \label{eq:RefinedGreatFilter}
Nrsq \not\gg 1 \, .
\end{equation}
Note that relations (\ref{eq:GreatFilter}) and (\ref{eq:RefinedGreatFilter})
are neither very precise statements, nor known truths, but should be thought
of as plausible conjectures given Fermi's great silence. 

For the Bayesian analysis, we think of $N$ as fixed and very large, and
need to devise a prior on $[0,1]^3$ for the three unknown parameters
$r$, $s$ and $q$. 
Estimating how likely emergence of life is on a typical planet, as
quantified by $r$ and $s$, is yet a basic unresolved question possibly
spanning physics, chemistry and biology (these could also inform us
about crucial barriers in $q$). Given that a Great Filter could
also relate to some difficulty of developing social complexity, 
social sciences may have a relevant role to play as well. 
Biogenesis suggests that $r>0$, but it could be very small (and
similarly for $s$). Without more precise knowledge, it makes sense to
have a prior that is spread out all over [0,1] for each parameter. A
natural first idea is to take the prior to be uniform on $[0,1]^3$,
corresponding to each of the three parameters being, independently of
the others, uniformly distributed on $[0,1]$. 

Our (partly hypothetical) 
data is as follows. First we have the great silence, which we represent
as $N$ independent Bernoulli trials with success
probability $rsq$, all taking value $0$ (failure to produce
noticeable super-technological civilisations). Then we consider the
effect of two different (hypothetical) possible observations. First,
to observe amoeba-level primitive life on a single planet, represented
as a Bernoulli trial with success probability $r$ taking value $1$
(success). Second, to observe a technological civilization on the
level of humanity today, a Bernoulli trial with success probability
$rs$. In particular, we are interested in what
effect this has on the statistical uncertainty about $q$.

At this point, we expect that attentive readers will suggest that we have
one more piece of data to take into account, namely the observation that
on planet Earth, we have evolved -- a probability $p$ ($=rs$) event. 
Following Bostrom (2008), 
we believe that including that piece of data in the Bayesian
conditioning would be a mistake, because no matter what the true value
of $p$ is, all observers would note that they themselves have evolved,
so our observation that we have evolved seems to offer no evidence one
way or the other about $p$. But the jury is still out on this somewhat
subtle issue, which seems to hinge on the choice between the so-called
self-sampling and self-indication assumptions in the study of observer
selection effects \citep{A11,H16,B02}.

If we accept the parameter $q$ as a guide to humanity's prospects of
making it all the way to becoming a super-technological civilization,
then the most relevant quantity to look at is the expected value of
$q$ under the posterior distribution. In particular, we are interested
in how that expected value changes when we switch from  the posterior
we get after just seeing the great silence to the one that also
incorporates the observation of primitive or human-level life on one
planet. 

\begin{figure}
\centering     
\includegraphics[width=140mm]{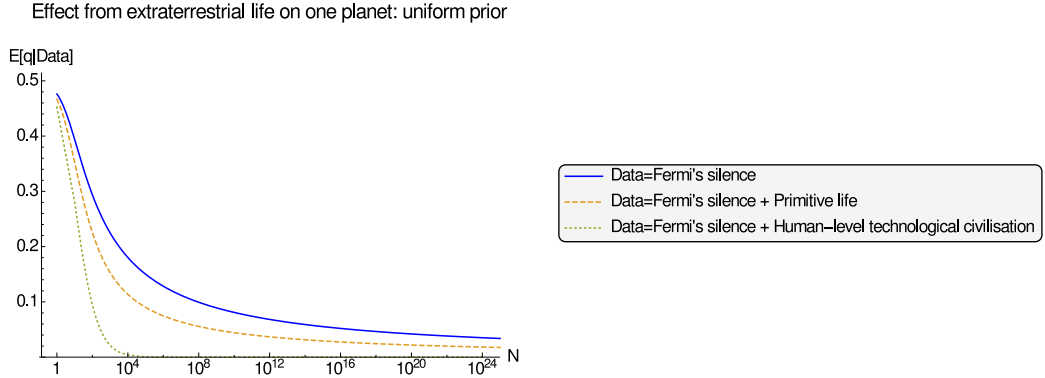}
\caption{\label{uniformPriorLargeN}The effect of extraterrestrial life
  on $q$ with a uniform prior for $r, s, q$ on $[0,1]^3$.}
\end{figure}

Figure \ref{uniformPriorLargeN} shows the effect of one observation of
primitive life, where we have taken the number of planets (experiments
described by parameters $r,s,q$) to vary. 
First, we see that only observing Fermi's silence gradually lowers the
expected value of $q$ with an increasing number of planets. This can
be understood from that the Bayesian posterior becomes proportional to
$(1-rsq)^N$, and with $N$ failed experiments getting bigger, it makes
the posterior reflect that one or more of $r$, $s$ and $q$ have to be
small. Second, we see the large effect that an additional observation
of primitive life has on the posterior, shifting the posterior belief
on $q$ much closer to zero. This can be understood as the new
observation shifting probability in the posterior towards larger
values of $r$,
and taken together with the observations from Fermi's silence
restricting $rsq$ to be small, this has to lower the other parameters
which decreases the expected value. The same line of reasoning holds
for observing human-level life, with belief shifted proportional to
$rs$, making the effect on $q$ much more drastic. The results are in line
with what we could expect from the preceding discussion, and shows one
way to quantify the Great Filter argument. Analogous calculations can
be made for the cases of the discoveries of  extinct primitive life (a
probability $r(1-s)$ event) and an extinct human-level civilization (a
probability $rs(1-q)$ event). 

But this is all for one particular choice of prior: uniform distribution
on $[0,1]^3$. As one of us has argued earlier \citep{H07}, uniform 
distribution is not in general a hallmark of objectivity, but just a model
assumption among many other possibilities, whence unreflected
and perfunctory use of it is bad practice. In this case, it may be argued 
that uniform distribution attaches unreasonably small probabilities
to very small values of the parameters. For instance, it is not a priori
implausible to imagine a universe in which biogenesis is possible 
(as it clearly is, or else we would not exist) but requires some very
low probability event, such as the chance assembly of some specific
and rather long RNA molecule (allowing biological evolution which
depends on replication to take over and make rapid progress by
becoming the dominating driver). This
could easily imply a value of $r$ less than, say, $10^{-50}$, but
uniform distribution assigns a mere $10^{-50}$ probability to the
event that $r$ takes such a small value, which is just too small to 
reflect our taking the possibility seriously. A distribution
that takes the possibility of very small parameter values more reasonably 
into account is the log-uniform, which was suggested in a similar setting
by Tegmark \citep{T14}. A log-uniform distribution on $[0,1]$ has
probability density proportional to $\frac{1}{x}$. Unfortunately, 
$\frac{1}{x}$ blows up near $x=0$ in such a way that 
$\int_0^1\frac{1}{x}dx=\infty$, so we have to truncate the distribution 
near $0$. Somewhat arbitrarily, we take the cutoff to be $10^{-100}$, 
giving $r$ density $\frac{1}{Cx}$ on $[10^{-100}, 1]$, where
$C= \int_{10^{-100}}^0\frac{1}{x}dx=100\ln(10)$ is a normalizing
constant to make it a probability measure. Independently, we let
$s$ and $q$ have the same distribution, so that the full prior is
concentrated on $[10^{-100}, 1]^3$ with density $\frac{1}{C^3xyz}$. 

\begin{figure}
\centering     
\includegraphics[width=140mm]{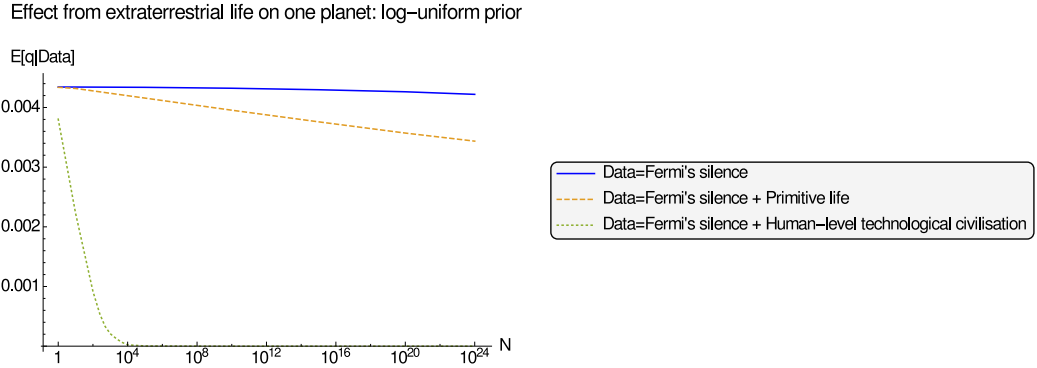}
\caption{\label{loguniformPriorLargeN}
The effect of extraterrestrial life on $q$ with a log-uniform prior
for $r, s, q$ in $[0,1]^3$}
\end{figure}

As far as the influence of discovering extraterrestrial life
on the expected value of $q$, this prior gives qualitatively similar results 
as the uniform prior: 
in Figure \ref{loguniformPriorLargeN}, we see that, similarly to Figure
\ref{uniformPriorLargeN}, observing primitive life always lowers the
expected value of $q$. Again, the effect on $q$ by an
observation of a human-level civilization is more drastic. Starting
out from different absolute levels of expected value reflects that we
start out from the prior with much larger weight on small
values. However, the remaining reasoning goes the same as the model of
failures in the Great Filter and success on one planet has a similar
effect with the new prior.

\begin{figure}
\centering     
\includegraphics[width=140mm]{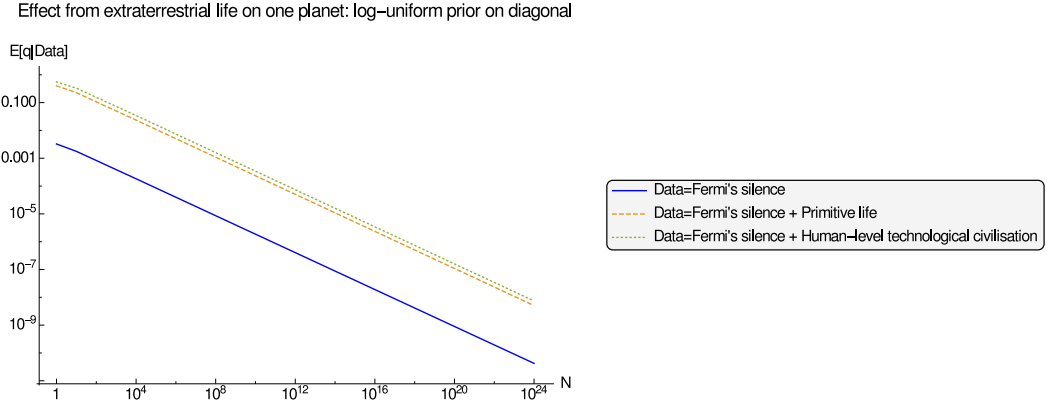}
\caption{\label{loguniformDIAGONALPriorLargeN}
The effect of extraterrestrial life on $q$ with a log-uniform prior
for $r, s, q$ on the diagonal $L=\{(x,y,z)\in [10^{-100}, 1]^3: x=y=z\}$.}
\end{figure}

In the two priors tried so far, the parameters $r$, $s$ and $q$ have
been taken to be independent. It is not clear that this is the most
reasonable choice to make. One way to reason is to suspect that all or
most universes in some (real or hypothetical) multiverse are either
generally amenable to producing complexity, or generally more sterile
(much like what seems to be the case with,  e.g., cellular automata
\citep{Wolfram}). In the former case, all of $r$, $s$ and $q$ can be
expected to be reasonably large, while in the latter case they can all
be expected to be near $0$. This suggests that in the prior, the
parameters should be positively correlated. For our next prior, we
shall consider the extreme case where they are all lined up on the
diagonal  $L=\{(x,y,z)\in [10^{-100}, 1]^3: x=y=z\}$, each
log-uniformly distributed as above, but no longer independent of each
other. This time, the result of discovering extraterrestrial life will
be rather different: in Figure \ref{loguniformDIAGONALPriorLargeN}, we
see that observing extraterrestial life has a positive effect on our
expectation of $q$. This can be understood since the parameters are
correlated: one instance of life on a lower level of complexity would
have the statistical effect to increase our estimates of $s$ and $q$
as well. 

A fair criticism against this last prior is that it dogmatically
confines the triple $(r,s,q)$ to lie on the diagonal $L$, thus
leaving huge parts of 
the space $[0,1]^3$ with zero probability. In the Appendix, 
we show that the example can be modified in such a way that the prior
is dense all over $[0,1]^3$ while preserving the same qualitative behavior
as regards the effect of discovering extraterrestrial life
on the expected value of $q$. 
Thus, we have quantitatively demonstrated that what we learn about $q$
and in which direction belief shifts after observing extraterrestrial
life can depend on which prior we use. This in turn highlights the
problem of understanding to which degree the universe is in general
amenable to produce complexity on various levels. Findings for or
against correlation between levels, such as theoretical results in
multiverse models \citep{T14}, could have an effect on which conclusion
results from the Great Filter argument.

\section{Conclusions}
In summary, we still think that the intuition about the alarming effect
of discovering extraterrestrial life expressed by Hanson (1998) and
Bostrom (2008) has some appeal. In our Bayesian analysis, our
first two priors (independent uniform, and independent log-uniform)
support it. The third one (perfectly correlated log-uniform), however, 
contradicts it, and while we find the prior a bit too extreme to make
a very good choice, this shows that some condition on the prior is
needed to obtain qualitative conclusions about the effect on $q$ of
discovering extraterrestrial life.

A final word of caution: While a healthy dose of critical thinking
regarding the choice of Bayesian prior is always to be recommended,
the case for epistemic humility is especially strong in the study of
the Fermi paradox and related ``big questions''. In more mainstream
scientific studies, circumstances are often favorable, either through
the existence of a solid body of independent evidence in support of
the prior, or through the availability of sufficient amounts of data
that one can reasonably hope that the effects of the prior are
(mostly) washed out in the posterior. In the present setting we have
neither, so all conclusions from the posterior should be viewed as
highly tentative.

\medskip\noindent
{\bf Acknowledgement.} We thank Thomas Ericsson for helpful comments
on numerical computation of the posterior. 

\bibliographystyle{agsm}

\bigskip\noindent
{\Large \bf Appendix}

\medskip\noindent
To demonstrate our claim that the expected value of $q$ can increase
in response to discovering extraterrestrials also in
cases where the prior is dense
in $[0,1]^3$, we need some terminology. Fix $N$. We write $\Pb_0$ for a 
probability measure corresponding to picking $(r,s,q)$ on the diagonal
$L$ as in our log-uniform diagonal example, plus doing the appropriate
Bernoulli trials for $N$ independent planets. We write $\Pb_1$ for 
the analogous thing, but with picking $(r,s,q)$ according to uniform
distribution on $[0,1]^3$. For $\varepsilon>0$, we let $\Pb_\varepsilon$
be the convex combination $\varepsilon\Pb_1 + (1-\varepsilon)\Pb_0$, and note
that since the distribution of $(r,s,q)$ is dense
in $[0,1]^3$ under $\Pb_1$,
the same is true under $\Pb_\varepsilon$. Write $\E_0$, $\E_1$ and $\E_\varepsilon$
for expectation under respectively $\Pb_0$, $\Pb_1$ and 
$\Pb_\varepsilon$.

Let $A$ be any event satisfying 
\begin{equation}  \label{eq:general_A}
\Pb_0(A)>0 \, . 
\end{equation}
Two examples of such $A$ is $A'$, defined as observing the great silence
($N$ independent Bernoulli ($rsq$) variables all taking value $0$), and
$A''$, defined as $A'$ plus discovering primitive life on a single planet. 
We have seen that
\begin{equation}  \label{eq:the_zero_case}
\E_0[q|A'] < \E_0[q|A'']
\end{equation}
and wish to show that
\begin{equation}  \label{eq:the_epsilon_case}
\E_\varepsilon[q|A'] < \E_\varepsilon[q|A'']
\end{equation}
for sufficiently small $\varepsilon>0$. To get from (\ref{eq:the_zero_case})
to (\ref{eq:the_epsilon_case}) it suffices to show that
\begin{equation}  \label{eq:the_needed_limit}
\lim_{\varepsilon \rightarrow 0} \E_\varepsilon[q|A] = \E_0[q|A]
\end{equation}
for all $A$ satisfying (\ref{eq:general_A}).
Writing $B$ for the event that $(r,s,q)$ sits on the diagonal
$r=s=q$, and decomposing $\E_\varepsilon[q|A]$
as
\[
\E_\varepsilon[q|A] = \Pb_\varepsilon(B|A)\E_\varepsilon[q|A,B]+
\Pb_\varepsilon(\neg B|A)\E_\varepsilon[q|A, \neg B]
\]
we first note that conditioning $\Pb_\varepsilon$ on the event $B$ yields 
$\Pb_0$, and that conditioning $\Pb_\varepsilon$ on the event $\neg B$ yields 
$\Pb_1$, so that
\begin{equation} \label{eq:decomposition}
\E_\varepsilon[q|A] = \Pb_\varepsilon(B|A)\E_0[q|A]+
\Pb_\varepsilon(\neg B|A)\E_1[q|A] \, . 
\end{equation}
As to the factor $\Pb_\varepsilon(B|A)$, we get using Bayes' Theorem that
\begin{eqnarray*}
\Pb_\varepsilon(B|A) 
& = & \frac{\Pb_\varepsilon(B)\Pb_\varepsilon(A|B)}{\Pb_\varepsilon(B)\Pb_\varepsilon(A|B)
+\Pb_\varepsilon(\neg B)\Pb_\varepsilon(A|\neg B) } \\
& = & \frac{(1-\varepsilon)\Pb_\varepsilon(A|B)}{(1-\varepsilon)\Pb_\varepsilon(A|B)
+\varepsilon\Pb_\varepsilon(A|\neg B) } \\
& = & \frac{(1-\varepsilon)\Pb_0(A)}{(1-\varepsilon)\Pb_0(A)
+\varepsilon\Pb_1(A) }
\end{eqnarray*}
which tends to $1$ as $\varepsilon \rightarrow 0$. It follows also that
$\lim_{\varepsilon\rightarrow 0} \Pb_\varepsilon(\neg B|A)=0$. Sending 
$\varepsilon\rightarrow 0$ in the right hand side of (\ref{eq:decomposition})
therefore yields (\ref{eq:the_needed_limit}), and
(\ref{eq:the_epsilon_case}) follows as desired.

\end{document}